\documentclass[conference]{IEEEtran}
\usepackage{amssymb}
\usepackage[dvips]{graphicx,color}
\newcommand{\qed}{\fbox{}}
\newcommand{\defeq}{\stackrel{\triangle}{=}}

\newtheorem{definition}{Definition}

\newtheorem{lemma}{Lemma}

\newtheorem{theorem}{Theorem}

\begin{document}
\title{On Random Construction of a Bipolar Sensing Matrix with Compact Representation} 
\author{
\IEEEauthorblockN{Tadashi Wadayama}
\IEEEauthorblockA{Nagoya Institute of Technology\\
Nagoya, Aichi, JAPAN\\
Email: wadayama@nitech.ac.jp}
}

\maketitle

\begin{abstract}
A random construction of bipolar sensing matrices based on 
binary linear codes is introduced and its RIP (Restricted Isometry Property) is analyzed
based on an argument on the ensemble average of the weight distribution of binary linear codes.
\end{abstract}

\section{Introduction}

Research in compressed sensing \cite{robust} \cite{LPdecoding} is expanding rapidly.
The sufficient condition for $\ell_1$-recovery  based on the 
Restricted Isometry Property (RIP) \cite{LPdecoding} \cite{RIP}  is one of the celebrated 
results in this field. The design of sensing matrices with small RIP constants is a theoretically 
interesting and challenging problem. Currently,  random constructions provide the strongest results, 
and the analysis of random constructions is based on large deviations 
of maximum and minimum singular values of random matrices  \cite {universal} \cite{LPdecoding}. 

In the present  paper, a random construction of bipolar sensing matrices based on 
binary linear codes is introduced and its RIP is analyzed.
The column vectors of the proposed sensing matrix
are nonzero codewords of a randomly chosen binary linear code. Using a generator matrix, a $p \times m$ sensing matrix 
can be represented by $O(p \log_2 m )$-bits. 
The existence of sensing matrices with the RIP is shown based on an argument on the ensemble average of the weight distribution of binary linear codes.

\section{Preliminaries}

\subsection{Notation}
The symbols $\Bbb R$ and $\Bbb F_2$ represent 
the field of real numbers and the finite field with two elements $\{0,1\}$, respectively.
The set of all $p \times m$ real matrices is denoted by $\Bbb R^{p \times m}$.
In the present paper,  the notation $x \in \Bbb R^p$ indicates that $x$ is a column vector of 
length $p$.
The notation $|| \cdot ||_p$ denotes $\ell_p$-norm $(1 \le p < \infty)$ defined by
\begin{equation}
||x||_p \defeq \left(\sum_{i=1}^p |x_i|^p \right)^{1/p}.
\end{equation}
The $\ell_0$-norm is defined by
\begin{equation}
||x||_0 \defeq |\mbox{supp(x)}|,
\end{equation}
where $\mbox{supp(x)}$ denotes the index set of nonzero components of $x$.
The functions $w_h(\cdot)$ and $d_h(\cdot, \cdot)$ are the Hamming weight and Hamming distance 
functions, respectively.

\subsection{Restricted isometry property (RIP)}
Let $\Phi \defeq \{\phi_1, \ldots,  \phi_m\}\in \Bbb R^{p \times m}$ be a $p \times m$ real matrix, where
the $\ell_2$-norm of the $j$-th $(j \in [1,m])$ column vector $\phi_j$ is normalized to 
one, namely, $||\phi_i||_2 = 1$. The notation $[a,b]$ represents the set of consecutive integers from $a$ to $b$.

The restricted isometry property of $\Phi$ introduced by Candes and Tao \cite{LPdecoding}
plays a key role in a sufficient condition of $\ell_1$-recovery.
\begin{definition}
A vector $x \in \Bbb R^m$ is called an $S$-sparse $(S \in  [1,m])$ vector 
if $||x||_0 \le S$.
If there exists a real number $\delta (0 \le \delta < 1)$ satisfying 
\begin{equation} \label{RIP}
(1-\delta) ||x||_2^2 \le ||\Phi x||_2^2 \le (1+\delta)||x||_2^2
\end{equation}
for any $S$-sparse vector $x \in \Bbb R^m$, then we say that $\Phi$ has the RIP of order $S$.
If $\Phi$ has the RIP of order $S$,  then the smallest constant satisfying (\ref{RIP})
is called the {\em RIP constant} of $\Phi$, which is denoted by $\delta_S$.
\hfill\qed
\end{definition}
Assume that $\Phi$ has the RIP with small $\delta_S$.
In such a case, any sub-matrix composed from $Q$-columns $(1 \le Q \le S)$ of $\Phi$
is nearly orthonormal.

Recently, Candes \cite{RIP} reported the relation between the RIP and the
$\ell_1$-recovery property. A portion of the main results of \cite{RIP} is summarized 
as follows. Let $S \in [1,m]$, and assume that $\Phi$ has the RIP with 
\begin{equation}\label{sqr2}
\delta_{2S} \le \sqrt{2}-1.
\end{equation}
For any $S$-sparse vector $e \in \Bbb R^m$ (i.e., $||e||_0 \le S$),
the solution of the following $\ell_1$-minimization problem 
\begin{equation}\label{l1recovery}
\mbox{minimize} ||d||_1 \mbox{ subject to } \Phi d = s
\end{equation}
coincides exactly with $e$, where $s = \Phi e$.
Note that \cite{RIP} considers stronger 
reconstruction results (i.e., robust reconstruction).
The matrix $\Phi$ in (\ref{l1recovery}) is called a {\em sensing matrix}.

\subsection{Relation between incoherence and the RIP}

The incoherence of $\Phi$ defined below and the RIP constant are closely related.
\begin{definition}
The incoherence of $\Phi$ is defined by 
\begin{equation}
\mu(\Phi) \defeq \max_{i, j \in [1,m], i \ne j} |\phi_i^T \phi_j|.
\end{equation}
\hfill\qed
\end{definition}

The following lemma shows the relation between the incoherence and the RIP constant.
Similar bounds are well known (e.g.,\cite{incoherence}).
\begin{lemma} \label{deltaupper}
Assume that $\Phi \in \Bbb R^{p \times m}$ is given.
For any $S \in [1,m]$, $\delta_S$ is upper bounded by
\begin{equation}
\delta_S <  \mu(\Phi) S.
\end{equation}
An elementary proof (different from that in \cite{incoherence}) is presented in Appendix.
\hfill\qed
\end{lemma}

\section{Construction of sensing matrices based on binary linear codes}

In this section, we present a construction method for sensing matrices based on 
binary linear codes.  A sensing matrix obtained from this construction has a
concise description. A sensor can store a generator matrix of a binary linear code, 
instead of the entire sensing matrix.

\subsection{Binary to bipolar conversion function}

The function $\beta_p: F_2^p \rightarrow \Bbb R^{p}$ is called a {\em binary to bipolar conversion function} 
defined by
\begin{equation}
\beta: x \in \Bbb F_2^p \mapsto \frac{1}{\sqrt{p}} (e - 2 x) \in \Bbb R^{p},
\end{equation}
where $e$ is an all-one column vector of length $p$. Namely, using the binary to bipolar conversion function,
a binary sequence is converted to a $\{+1/\sqrt{p}, -1/\sqrt{p} \}$-sequence.

The following lemma demonstrates that the inner product of two bipolar sequences $\beta_p(a)$ and $\beta_p(b)$ is determined from 
the Hamming distance between the binary sequences $a$ and $b$.

\begin{lemma}\label{hammingandcorr}
For any $a, b \in \Bbb F_2^p$, the inner product of $\beta_p(a)$ and $\beta_p(b)$ is given by
\begin{equation} \label{correqu}
\beta_p(a)^T  \beta_p(b) = 1 - \frac{2 d_h(a,b) }{p}.
\end{equation}
(Proof) Let $\beta_p(a) = (a_1,\ldots, a_p)^T$ and $\beta_p(b)= (b_1,\ldots, b_p)^T$.
Define $Y_{1}$ and $Y_{2}$ by
\begin{equation}
Y_{1} \defeq \{i \in [1,p]: a_i = b_i \}, \quad Y_{2} \defeq \{i \in [1,p]: a_i \ne b_i \},
\end{equation}
where $|Y_1| = p - d_h(a,b)$ and $|Y_2| =  d_h(a,b)$.  Equation (\ref{correqu}) 
is derived as follows:
\begin{eqnarray} \nonumber
\beta_p(a)^T  \beta_p(b)  
&=& \sum_{i=1}^p a_i b_i \\ \nonumber
&=& \sum_{i \in Y_1}^p a_i b_i  + \sum_{i \in Y_2}^p a_i b_i   \\ \nonumber
&=& \sum_{i \in Y_1}^p \frac{1}{p}  + \sum_{i \in Y_2}^p \left(-\frac{1}{p} \right)    \\ \nonumber
&=& (p - d_h(a,b)) \frac{1}{p}  + d_h(a,b) \left(-\frac{1}{p} \right)    \\ 
&=& 1 - \frac{2 d_h(a,b) }{p}.
\end{eqnarray}
\hfill\qed
\end{lemma}
It is easy to confirm that $\beta_p(a)$ is normalized, i.e., $ ||\beta_p(a)||_2 = 1$,
for any $a \in \Bbb F_2^p$.

\subsection{Construction of the sensing matrix}

Let $H \in \Bbb F_2^{r \times p}$ $(p > r)$ be a 
binary $r \times p$ parity check matrix where $2^{p-r} \ge p$ holds.
The binary linear coded $C(H)$ defined by
$H$ is given by
\begin{equation}
C(H) \defeq \{x \in \Bbb F_2^p: H x = 0^r \},
\end{equation}
where $0^r$ is a zero-column vector of length $r$.
The following definition gives the construction of sensing matrices.
\begin{definition}
Assume that all of the nonzero 
codewords of $C(H)$ are denoted by $c_1,c_2,\ldots, c_M$ (based on any predefined order),
where $M = 2^{p-rank(H)} - 1 \ge 2^{p-r} -1$. 
The sensing matrix $\Phi(H) \in \Bbb R^{p \times m}$ is defined by
\begin{equation}
\Phi(H) \defeq \left(\beta_p(c_1),\beta_p(c_2), \ldots, \beta_p(c_m) \right),
\end{equation}
where $m = 2^{p-r}-1$.
If $\Phi(H)$ has the RIP of order $S$, the RIP constant corresponding to $\Phi(H)$ is denoted by $\delta_S(H)$.
\hfill\qed
\end{definition}
Since the order of the columns is unimportant,
we do not distinguish between sensing matrices of different column order (or choice of codewords from $C(H)$). 

If the weights of all nonzero codewords of $C(H)$ are very close to $p/2$, 
then the incoherence of $\Phi(H)$ becomes small, as described in detail in the following lemma.
\begin{lemma} \label{RIPbound}
Assume that $\epsilon (0 < \epsilon < 1)$ is given and that 
\begin{equation}
\left(\frac{1-\epsilon}{2} \right)p \le w_h(c) \le \left(\frac{1+\epsilon}{2} \right)p
\end{equation}
holds for any $c \in C(H)\backslash 0^p$.
In such a case, the incoherence $\Phi(H)$ is 
upper bounded by
\begin{equation}
\mu(\Phi(H)) \le \epsilon.
\end{equation}
(Proof)  For any pair of codewords $a,b (a \ne b) \in C(H)$,  the Hamming weight of $a + b$ is
in the range:
\begin{equation}
\left(\frac{1-\epsilon}{2} \right)p \le w_h(a + b)  \le \left( \frac{1+\epsilon}{2} \right)p.
\end{equation}
due to the linearity of $C(H)$. This means that 
\begin{equation}
\left( \frac{1-\epsilon}{2} \right)p \le d_h(a, b)  \le \left(\frac{1+\epsilon}{2} \right)p
\end{equation}
holds for any $a,b \in C(H) (a \ne b)$. Using Lemma \ref{hammingandcorr}, we immediately obtain
\begin{equation}\label{corrange}
\forall i,j (i \ne j) \in [1,m],\quad -\epsilon \le \beta_p(c_i)^T  \beta_p(c_j) \le \epsilon,
\end{equation}
where
\begin{equation}
\Phi(H) = \left(\beta_p(c_1),\beta_p(c_2), \ldots, \beta_p(c_m) \right).
\end{equation}
The definition of incoherence and the above inequalities lead to an upper bound on 
the incoherence: 
\begin{equation}
\mu(\Phi(H)) \le \epsilon.
\end{equation}
\hfill\qed
\end{lemma}

\subsection{Analysis based on ensemble average of weight distribution}

We here consider binary linear codes 
whose weight distribution is tightly concentrated around the Hamming weight $p/2$. 
Before starting the analysis, we introduce 
the weight distribution $\{A_w(H)\}_{w\in [1,n]}$, which is defined by 
\begin{equation}
A_w(H) \defeq |\{c: c \in C(H), w_h(c) = w  \}|.
\end{equation}
In the present paper, we consider an ensemble of 
binary parity check matrices, which is referred to herein as the {\em random ensemble}.
The random ensemble $R_{r,p}$ contains all binary $r \times p$ matrices.
Equal probability 
$
P(H) = 1/2^{rp}
$
is assigned to each matrix in $R_{r,p}$.
Let $f$ be a real-valued function defined on $R_{r,p}$, which can be 
considered as a {\em random variable} defined over the ensemble $R_{r,p}$.
The expectation of $f$ with respect to the ensemble $R_{r,p}$ is defined by
\begin{equation}
E_{R_{r,p}} [f] \defeq \sum_{H \in R_{r,p}} P(H) f(H).
\end{equation}
The expectation of weight distributions with respect to the random ensemble has been reported \cite{Gal63} to be 
\begin{equation}
E_{R_{r,p}}[A_w(H)] = {p \choose w} 2^{-r}.
\end{equation}

In the following, a combination of average weight distribution and Markov inequality is used to show that the RIP holds for $\Phi(H)$ with overwhelmingly high probability.
\begin{lemma} \label{first}
Assume that we draw a parity check matrix from $R_{r,p}$.
The probability of selecting $H$ that satisfies $\mu(\Phi(H)) \le \epsilon$  is lower bounded by 
\begin{equation}
1 - 2^{1-r}\sum_{w=0}^{\lfloor (\frac{1-\epsilon}{2} )p \rfloor}  {p \choose w } .
\end{equation}
(Proof)
Let us define $K_\epsilon(H)$ as 
\begin{equation}
K_\epsilon(H) \defeq
\sum_{w=1}^{\lfloor (\frac{1-\epsilon}{2} )p \rfloor} A_w(H)
+ \sum_{w=\lceil (\frac{1+\epsilon}{2} )p \rceil}^{p} A_w(H)
\end{equation}
for $H \in R_{r,p}$. 
The condition  $K_\epsilon(H)=0$ implies that
\begin{equation}
\left( \frac{1-\epsilon}{2} \right)p \le w_h(c) \le \left(\frac{1+\epsilon}{2} \right)p
\end{equation}
for any $c \in C(H) \backslash 0^p$.
Namely, if $K_\epsilon(H)=0$ holds, then $\mu(\Phi(H))$ is proven to be smaller than or equal to $\epsilon$ 
by Lemma \ref{RIPbound}. Next, we evaluate the ensemble expectation of $K_\epsilon(H)$:
\begin{eqnarray} \nonumber
E_{R_{r,p}}[K_\epsilon(H)] 
&=&
\sum_{w=1}^{\lfloor (\frac{1-\epsilon}{2} )p \rfloor} E_{R_{r,p}}[A_w(H)] \\ \nonumber
&+& \sum_{w=\lceil (\frac{1+\epsilon}{2} )p \rceil}^{p} E_{R_{r,p}}[A_w(H)] \\ \nonumber
&=&
\sum_{w=1}^{\lfloor (\frac{1-\epsilon}{2} )p \rfloor} 2^{-r} {p \choose w } 
+ \sum_{w=\lceil (\frac{1+\epsilon}{2} )p \rceil}^{p} 2^{-r} {p \choose w } \\
&<&
2^{1-r} \sum_{w=0}^{\lfloor (\frac{1-\epsilon}{2} )p \rfloor}  {p \choose w }.
\end{eqnarray}
The final inequality is due to the following identity on the binomial coefficients:
\begin{equation}
\forall w \in [0,p],\quad {p \choose w} = {p \choose p - w }.
\end{equation}
Using the Markov inequality, we obtain the following upper bound on the probability 
of the event  $K_\epsilon(H) \ge 1$:
\begin{eqnarray} \nonumber
Prob[K_\epsilon(H) \ge 1] 
&\le& E_{R_{r,p}}[K_\epsilon(H)]  \\ 
&<& 2^{1-r} \sum_{w=0}^{\lfloor (\frac{1-\epsilon}{2} )p \rfloor}  {p \choose w } .
\end{eqnarray}
Since $K_{\epsilon}(H)$ takes a non-negative integer-value,  we have 
\begin{equation}
Prob[K_\epsilon(H) = 0] 
> 1 - 2^{1-r} \sum_{w=0}^{\lfloor (\frac{1-\epsilon}{2} )p \rfloor}  {p \choose w }.
\end{equation}
This completes the proof.
\hfill\qed
\end{lemma}

The following theorem is the main contribution of the present paper.
\begin{theorem} \label{main}
Assume that $H$ is chosen randomly according to the probability assignment of
$R_{r,p}$. If 
\begin{equation}
S   < Z \sqrt{\frac{p}{\log_2 m}},
\end{equation}
holds, then $\delta_{2S}(H) < \sqrt{2}-1$ holds with probability greater than 
\begin{equation}
1 - 2^{1 - p + r},
\end{equation}
where $m = 2^{p-r}-1$.
The constant $Z$ is given by
\begin{equation}
Z \defeq \frac{\sqrt{2} -1 }{2 \sqrt{6}}.
\end{equation}
(Proof) A simpler upper bound on 
\begin{equation} \label{simpler}
2^{1-r}\sum_{w=0}^{\lfloor (\frac{1-\epsilon}{2} )p \rfloor}  {p \choose w }
\end{equation}
is required.
Using the inequality on binomial coefficients \cite{Cover}:
\begin{equation}
{p \choose w}  \le 2^{p H(w/p)},
\end{equation}
we have 
\begin{eqnarray} \nonumber
 2^{1-r}\sum_{w=0}^{\lfloor (\frac{1-\epsilon}{2} )p \rfloor } {p \choose w }
&\le&  
 2^{1-r}\sum_{w=0}^{\lfloor (\frac{1-\epsilon}{2} )p \rfloor } 2^{p H(w/p)} \\ \nonumber
&<&   2^{1-r} \times p \times 2^{p H\left(\frac{1-\epsilon}{2} \right)}  \\
&=&    2^{1-r  + \log_2 p + p H\left(\frac{1-\epsilon}{2} \right)}, 
\end{eqnarray}
where $H(x)$ is the binary entropy function defined by
\begin{equation}
H(x) \defeq -x \log_2 x -(1-x) \log_2 (1-x).
\end{equation}
In order to consider the exponent of an upper bound,
we take the logarithm of (\ref{simpler}) and obtain an upper bound of the exponent:
\begin{eqnarray} \nonumber
\log_2 \left[ 2^{1-r}\sum_{w=0}^{\lfloor (\frac{1-\epsilon}{2} )p \rfloor } {p \choose w } \right]
\hspace{-4mm}&<&\hspace{-2mm}1  + \log_2 (m+1) - p  + \log_2 p  \\ \label{exponentp}
&+& \hspace{-2mm}p H\left(\frac{1-\epsilon}{2} \right) \\  
&<&\hspace{-2mm}1  + 2\log_2 (m+1) - \frac{1}{2} p \epsilon^2.  
\end{eqnarray}
In the above derivation, we used the relation 
\begin{equation}
r = p - \log_2 (m+1)
\end{equation}
and the assumption $2^{p-r} \ge p$. 
A quadratic upper bound on the binary entropy function (Lemma \ref{entropybound} in Appendix) was also exploited
to bound the entropy term.

Letting 
\begin{equation}
\epsilon \defeq \sqrt{\frac{6\log_2 (m+1)}{p}},
\end{equation}
we have 
\begin{eqnarray} \nonumber
1  + 2\log_2 (m+1) - \frac{1}{2} p \epsilon^2 
&=& 1 - \log_2 (m+1) \\ 
&=& 1 - p + r.
\end{eqnarray}
Lemma \ref{deltaupper} and Lemma \ref{first} imply that, in this case, $\delta_S(H) < \epsilon S$ holds with probability greater than $1- 2^{1 - p + r}$.
Due to Lemma \ref{deltaupper}, the $\ell_1$-recovery condition (\ref{sqr2}) can be written as
\begin{eqnarray}
\delta_{2S } < 2  \sqrt{\frac{6\log_2 (m+1)}{p}} S < \sqrt{2} - 1.
\end{eqnarray}
From this inequality, we have
\begin{equation}
S   < Z \sqrt{\frac{p}{\log_2 (m+1) }} < Z \sqrt{\frac{p}{\log_2 m }} ,
\end{equation}
which proves the claim of the theorem.
\hfill\qed
\end{theorem}

\subsection{Asymptotic analysis}
In this subsection, the asymptotic properties of the proposed construction are given.
\begin{lemma} \label{second}
Assume that we draw a parity check matrix from $R_{r,p}$.
The probability of selecting $H$ that satisfies $\mu(\Phi(H)) \le \epsilon$  is upper bounded by 
\begin{equation} \label{secondmoment}
\frac{(1-2^{-r})2^{1+r}\sum_{w=0}^{\lfloor (\frac{1-\epsilon}{2} )p \rfloor}  {p \choose w } }
{ \left(2 \sum_{w=0}^{\lfloor (\frac{1-\epsilon}{2} )p \rfloor}  {p \choose w } -1   \right)^2}.
\end{equation}
(Proof) Here, we use a variant of Chebyschev's inequality \cite{alon}: 
\begin{equation} \label{Chebyschev}
Prob[K_{\epsilon}(H) = 0] \le \frac{VAR_{R_{r,p}}(K_{\epsilon}(H))}{E_{R_{r,p}}[K_{\epsilon}(H) ]^2},
\end{equation}
where $VAR_{R_{r,p}}(\cdot)$ denotes the variance with respect to $R_{r,p}$.
The variance $VAR_{R_{r,p}}(K_{\epsilon}(H))$ is given by
\begin{eqnarray} \nonumber
&&\hspace{-10mm}VAR_{R_{r,p}}(K_{\epsilon}(H)) \\ \nonumber
&=& \hspace{-4mm}\sum_{w_1=1}^A \sum_{w_2=1}^A Cov(w_1,w_2) 
+ \sum_{w_1=1}^A \sum_{w_2=B}^p  Cov(w_1,w_2) \\  \label{covcov}
&+& \hspace{-5mm}\sum_{w_1=B}^p \sum_{w_2=1}^A Cov(w_1,w_2) 
+ \sum_{w_1=B}^p \sum_{w_2=B}^p Cov(w_1,w_2),
\end{eqnarray}
where $A = \lfloor (1-\epsilon)p/2 \rfloor$ and $B = \lceil (1+\epsilon)p/2 \rceil$.
The covariance of weight distributions denoted by $Cov(w_1,w_2)$ is defined as follows:
\begin{eqnarray}  \nonumber
Cov(w_1,w_2) 
&\defeq& E_{R_{r,p}}[A_{w_1}(H)A_{w_2}(H) ]  \\
&-& E_{R_{r,p}}[A_{w_1}(H)] E_{R_{r,p}}[A_{w_2}(H) ]
\end{eqnarray}
for $w_1, w_2 \in [1,n]$. The covariance for the random ensemble 
has the following closed formula \cite{undetected}:
\begin{equation}
Cov(w_1,w_2) = 
\left\{
\begin{array}{ll}
(1-2^{-r})2^{-r} {p \choose w} & w_1 = w_2 = w \\
0   & w_1 \ne w_2
\end{array}
\right.
\end{equation}
for $w_1, w_2 \in [1,n]$. Applying the covariance formula to (\ref{covcov}), 
we have
\begin{eqnarray} \nonumber
&&\hspace{-8mm}VAR_{R_{r,p}}(K_{\epsilon}(H))  \\ \nonumber
&=& (1-2^{-r})2^{-r} \left( \sum_{w=1}^A{p \choose w} +  \sum_{w=B}^n {p \choose w}    \right) \\ \label{covbound}
&< & (1-2^{-r})2^{1-r}\sum_{w=0}^A{p \choose w}. 
\end{eqnarray}
Plugging the expectation of $K_\epsilon(H)$
\begin{eqnarray} \nonumber
E_{R_{r,p}}[K_\epsilon(H)] 
&=&
2^{-r}\left( \sum_{w=1}^{A}  {p \choose w } + \sum_{w=B}^{p} {p \choose w }  \right) \\
&=&
2^{-r}\left( 2\sum_{w=0}^{A}  {p \choose w } -1  \right) 
\end{eqnarray}
and the upper bound on the variance (\ref{covbound}) into (\ref{Chebyschev})
proves the lemma.
\hfill\qed
\end{lemma}

The asymptotic behavior of $Prob[K_\epsilon(H)=0]$ and $Prob[K_\epsilon(H)\ne0]$
is summarized  in the following theorem. 
\begin{theorem} \label{asymptoticth}
Assume that $\alpha = r/p$ is fixed $(0 < \alpha < 1)$.
Let 
\begin{eqnarray} 
f_1(\epsilon, \alpha) &\defeq& \lim_{p \rightarrow \infty} \frac 1 p \log_2 Prob[K_\epsilon(H)=0] \\
f_2(\epsilon, \alpha) &\defeq& \lim_{p \rightarrow \infty} \frac 1 p \log_2 Prob[K_\epsilon(H)\ne0].
\end{eqnarray}
The following inequalities give upper bounds on $f_1(\epsilon)$ and $f_2(\epsilon)$, respectively:
\begin{eqnarray} \label{f1}
f_1(\epsilon, \alpha) &<& \alpha - H\left(\frac{1-\epsilon}{2} \right), \\ \label{f2}
f_2(\epsilon, \alpha) &<& -\alpha + H\left(\frac{1-\epsilon}{2} \right).
\end{eqnarray}
(Proof) We first discuss (\ref{f1}). Let 
\begin{equation}
X \defeq \sum_{w=0}^{\lfloor (1-\epsilon)p/2  \rfloor} {p \choose w}.
\end{equation}
Using the inequality on the binomial coefficients
\begin{equation}
{p \choose w} \ge \frac{1}{(p+1)^2} 2^{p H(w/p)},
\end{equation}
$X$ can be  bounded from below:
\begin{eqnarray} \nonumber
X 
&>& {p \choose \lfloor (1-\epsilon)p/2 \rfloor } \\ \label{entropylower}
&\ge& \frac{1}{(p+1)^2}2^{p H\left(  (1-\epsilon)/2 -1/p   \right)}.
\end{eqnarray}
The inequality (\ref{secondmoment}) can be simplified as
\begin{eqnarray}
\frac{(1-2^{-\alpha p}) 2^{1+\alpha p}X  }{(2X-1)^2} < 2^{1+\alpha p} X^{-1}
\end{eqnarray}
for sufficiently large $X$. The right-hand side of the above inequality can be bounded from above 
using (\ref{entropylower}):
\begin{equation}
2^{1+\alpha p} X^{-1} \le (p+1)^2 2^{1+\alpha p -p H\left(  (1-\epsilon)/2 -1/p   \right)}.
\end{equation}
We are now able to derive the inequality given in (\ref{f1}) as follows:
\begin{eqnarray} \nonumber
&& \hspace{-12mm} \lim_{p \rightarrow \infty} \frac 1 p \log_2 \left[(p+1)^2 2^{1+\alpha p -p H\left(  (1-\epsilon)/2 -1/p   \right)} \right] \\
&=& \alpha - H\left(\frac{1-\epsilon}{2} \right).
\end{eqnarray}
The inequality given in (\ref{f2}) is readily obtained from (\ref{exponentp}).
\hfill\qed
\end{theorem}

Theorem \ref{asymptoticth} implies a sharp threshold behavior in the asymptotic regime.
Let $\alpha^*(\epsilon)$ be 
\begin{equation}
\alpha^*(\epsilon) \defeq H\left(\frac{1 - \epsilon}{2} \right),
\end{equation}
which is referred to as the {\em critical exponent}.
If $\alpha < \alpha^*(\epsilon)$, (\ref{f1}) means that the probability to draw a $p \times r$ matrix with $\mu(\Phi(H)) \le \epsilon$
decreases exponentially as $p$ goes to infinity.
On the other hand,  (\ref{f2}) indicates that 
the probability {\em not} to select a matrix with $\mu(\Phi(H)) \le \epsilon$
decreases exponentially if $\alpha > \alpha^*(\epsilon)$.

\section{Concluding remarks}

In the present paper, a construction of a bipolar sensing matrix is introduced and its RIP is analyzed. The existence of sensing matrices with the RIP has been shown based on  a probabilistic argument. An advantage of this type of sensing matrix is its compactness.
A sensor  requires $O(pm)$-bits in order to store a truly random $p \times m$ bipolar matrix.
On the other hand,  we need only $O(p \log_2 m)$-bits to store $\Phi(H)$
because we can use a generator matrix of $C(H)$ as a compact representation of $C(H)$.
However, this limited randomness of matrices results in a penalty on the RIP constant.
Although the present construction is based on a probabilistic construction, 
the results shown in Theorem \ref{main} are weaker than 
the $\ell_1$-recovery condition $O(S \log_e(m/S)) < p$ for the truly random $p \times m$ bipolar matrix ensemble shown in \cite{universal}. The condition shown in Theorem \ref{main} can be written as $O(S \sqrt{\log_2 m}) < \sqrt{p}$ and is more similar to the conditions of deterministic constructions, such as that given in \cite{DeVore}.

Lemma \ref{RIPbound} may be useful for evaluating the goodness of a randomly generated instance. The weight distribution of $C(H)$ can be evaluated with time complexity $O(m p)$, and an upper bound on the RIP constant can be obtained using Lemma \ref{RIPbound}.

\section*{Appendix}
\begin{lemma}\label{entropybound}
The following inequality\footnote{This bound becomes tighter as $x$ approaches to $1/2$.} 
\begin{equation}
-2\left(x- \frac 1 2 \right)^2 \ge H(x)-1
\end{equation}
holds for  $0 <  x < 1$. \\
(Proof)
Let $f(x)$ be 
\begin{equation}
f(x) \defeq -2\left(x- 1/2 \right)^2 -(H(x)-1)
\end{equation}
the domain of which is $0 < x < 1$.
The first and second derivatives of $f(x)$ are given by
\begin{equation}
f'(x) = -4\left(x- 1/2 \right) - \log_2 (1-x) + \log_2 x
\end{equation}
and
\begin{equation}
f''(x) = -4 +\left(\frac{1}{1-x}+\frac{1}{x}  \right) \frac{1}{\log_e 2},
\end{equation}
respectively. It is easy to verify that  $f''(x)>0$ for  $0 < x < 1$, which indicates that $f(x)$ is convex. Thus, we can obtain the global minimum of $f(x)$ by
solving $f'(x) = 0$, and we have $f'(1/2)=0$ and $f(1/2) = 0$. 
 \hfill\qed
\end{lemma}

\subsection*{Proof of Lemma \ref{deltaupper}}
Let $Q$ be an index set satisfying $Q \subset \{1,\ldots, m\}, |Q| \le  S$.
For any $c = (c_i)_{i \in Q} \in \Bbb R^{|Q|}$, we have 
\begin{eqnarray} \nonumber
||\Phi_Q c||_2^2 
&=& (\Phi_Q c)^T (\Phi_Q c) = \left(\sum_{i \in Q} c_i \phi_i  \right)^T \left(\sum_{j \in Q} c_j \phi_j  \right) \\ \nonumber
&=& \sum_{i \in Q} \sum_{j \in Q} c_i c_j \phi_i^T \phi_j 
= \sum_{i \in Q} c_i^2 + \sum_{i,j  \in Q (i \ne j)} c_i c_j \phi_i^T \phi_j \\ \nonumber
&\le& \sum_{i \in Q} c_i^2 + \sum_{i,j  \in Q (i \ne j)} |c_i c_j \phi_i^T \phi_j| \\ \label{cicj}
&\le& \sum_{i \in Q} c_i^2 +  \mu(\Phi) \sum_{i,j  \in Q (i \ne j)} |c_i c_j |,
\end{eqnarray}
where $\Phi_Q$ is a sub-matrix of $\phi$ composed from the columns corresponding to the index set $Q$.
For any $a, b \in \Bbb R$, 
\begin{equation}
(a^2 + b^2)/2  \ge |a b| \label{abstbound}
\end{equation}
holds since 
$
(|a| - |b|)^2 = a^2 + b^2 - 2 |a b| \ge 0.
$
We use this inequality to bound $|c_i c_j|$ in (\ref{cicj}) and obtain 
\begin{eqnarray} \nonumber
||\Phi_Q c||_2^2 
&\le& \sum_{i \in Q} c_i^2 +  \mu(\Phi) \sum_{i,j  \in Q (i \ne j)} |c_i c_j|  \\ \nonumber
&<& \sum_{i \in Q} c_i^2 +  \mu(\Phi) \sum_{i,j  \in Q } \left(\frac{c_i^2 + c_j^2}{2}\right)  \\ \nonumber
&=& \sum_{i \in Q} c_i^2 + \mu(\Phi)|Q| \sum_{i  \in Q } c_i^2   \\ \nonumber
&=& ||c||_2^2 (1  + \mu(\Phi)|Q| )\\ \nonumber
&\le& ||c||_2^2 (1  + \mu(\Phi)S ).
\end{eqnarray}
Similarly, $||\Phi_Q c||_2^2$ can be lower bounded by
$
||\Phi_Q c||_2^2 \ge ||c||_2^2 (1   - \mu(\Phi)S ).
$
From the definition of $\delta_S$, the lemma is proven. 

\section*{Acknowledgment}
The author would like to thank the anonymous reviewers of IEEE Information Theory Workshop  2009 
for their constructive comments.
The present study was supported in part by the Ministry of Education, Science, Sports
and Culture of Japan through a Grant-in-Aid for Scientific Research on Priority Areas
(Deepening and Expansion of Statistical Informatics) 180790091.

\end{document}